\documentclass[twocolumn,showpacs,preprintnumbers,amsmath,amssymb,
]
{revtex4}

\usepackage{graphicx}
\usepackage{dcolumn}
\usepackage{bm}

\def\bea{\begin{eqnarray}}
\def\eea{\end{eqnarray}}

\keywords{blah blah}

\begin{document}

\preprint{Version 0.1}

\title{Soft and hard components of two-particle distributions on $(y_t,\eta,\phi)$ \\ from p-p collisions at $\sqrt{s}$ = 200 GeV}
\thanks{STAR collaboration poster presented at Quark Matter 2004 by R.\,J.~Porter.}
\author{R.\,J.~Porter and T.\,A.~Trainor
}
\affiliation{STAR Collaboration}
\address{ University of Washington, Seattle, Washington, 98195}

\date{\today}

\begin{abstract}
We report measurements of large-scale two-particle correlations for 200 GeV p-p collisions on momentum components transverse rapidity $y_t$ (pion mass assigned), pseudorapidity $\eta$ and azimuth angle $\phi$.  In both transverse $y_t \otimes y_t$ and axial $(\eta\otimes\eta,\phi\otimes\phi)  $ two-particle subspaces we observe two components of correlation structure (soft and hard) which we interpret respectively in terms of longitudinal string fragmentation and transverse minimum-bias parton fragmentation (minijets). This combination is also represented by Lund-model-based Monte Carlo simulations such as Pythia.
\end{abstract}


\maketitle

\section{Introduction}

The two-component model of p-p collisions describes final-state hadrons as arising from a combination of longitudinal (along beam direction) string fragmentation (soft) described by the Lund string-fragmentation model~\cite{lund} and transverse parton fragmentation (hard) described by fragmentation functions derived from deep-inelastic scattering~\cite{frag}. In this context we have studied two-particle correlations for unidentified hadrons from 200 GeV p-p collisions observed with the STAR detector at RHIC~\cite{jeffp,jeffqm04}. We find that the two components are easily separated by means of a cut space defined on transverse rapidity. Simple model functions precisely characterize the data and represent all correlation structure with a few parameters. Comparison is made to a similar analysis of Pythia~\cite{pythia} p-p collisions. 

\section{Analysis method}

Single-particle momentum space for high-energy nuclear collisions can be represented by kinematic variables $(y_t,\eta,\phi)$ relative to the collision axis, where $p_t$ is transverse momentum, $m_t \equiv \sqrt{p_t^2 + m_\pi^2}$ is transverse mass, $y_t \equiv \ln\{(m_t + p_t)/m_\pi\}$ is transverse rapidity assuming pion mass $m_\pi$ for all particles, $\eta$ is the pseudorapidity, and $\phi$ is the momentum azimuth angle. The full two-particle momentum space $\vec{p} \otimes \vec{p}$ can be decomposed into subspaces represented by Cartesian products such as transverse-rapidity space $y_t \otimes y_t$. That space, in addition to revealing its own interesting correlation structure, serves as a cut space to isolate soft and hard components for further study in complementary {\em axial} momentum subspaces $\eta \otimes \eta$ and $\phi \otimes \phi$. 

Pair-number ratio histograms $\hat r_{ij} = \hat n_{ij,sib} / \hat n_{ij,mix}$, where $\hat n$ is a normalized pair-number histogram for sibling (same event) or mixed (different events) pairs, are formed on each two-particle momentum subspace. Pair-ratio distributions on $\eta \otimes \eta$ and $\phi \otimes \phi$ can be projected as {\em autocorrelations} onto individual difference variables $\eta_\Delta = \eta_1 - \eta_2$ and $\phi_\Delta = \phi_1 - \phi_2$, and onto their Cartesian product $\eta_\Delta \otimes \phi_\Delta$ to form a {\em joint} autocorrelation. Plotted in the figures is differential quantity $\bar N (\hat r - 1)$ ($\bar N$ is the ensemble mean total multiplicity within the detector acceptance) which measures correlations per final-state particle. Pair-ratio distributions are determined for like-sign (LS) and unlike-sign (US) charge combinations and for charge-independent (isoscalar, CI = LS $+$ US) and charge-dependent (isovector, CD = LS $-$ US) combinations.

\section{$y_t \otimes y_t$ as cut space}

\begin{figure}[h]
\includegraphics[height=1.5in,width=1.5in]{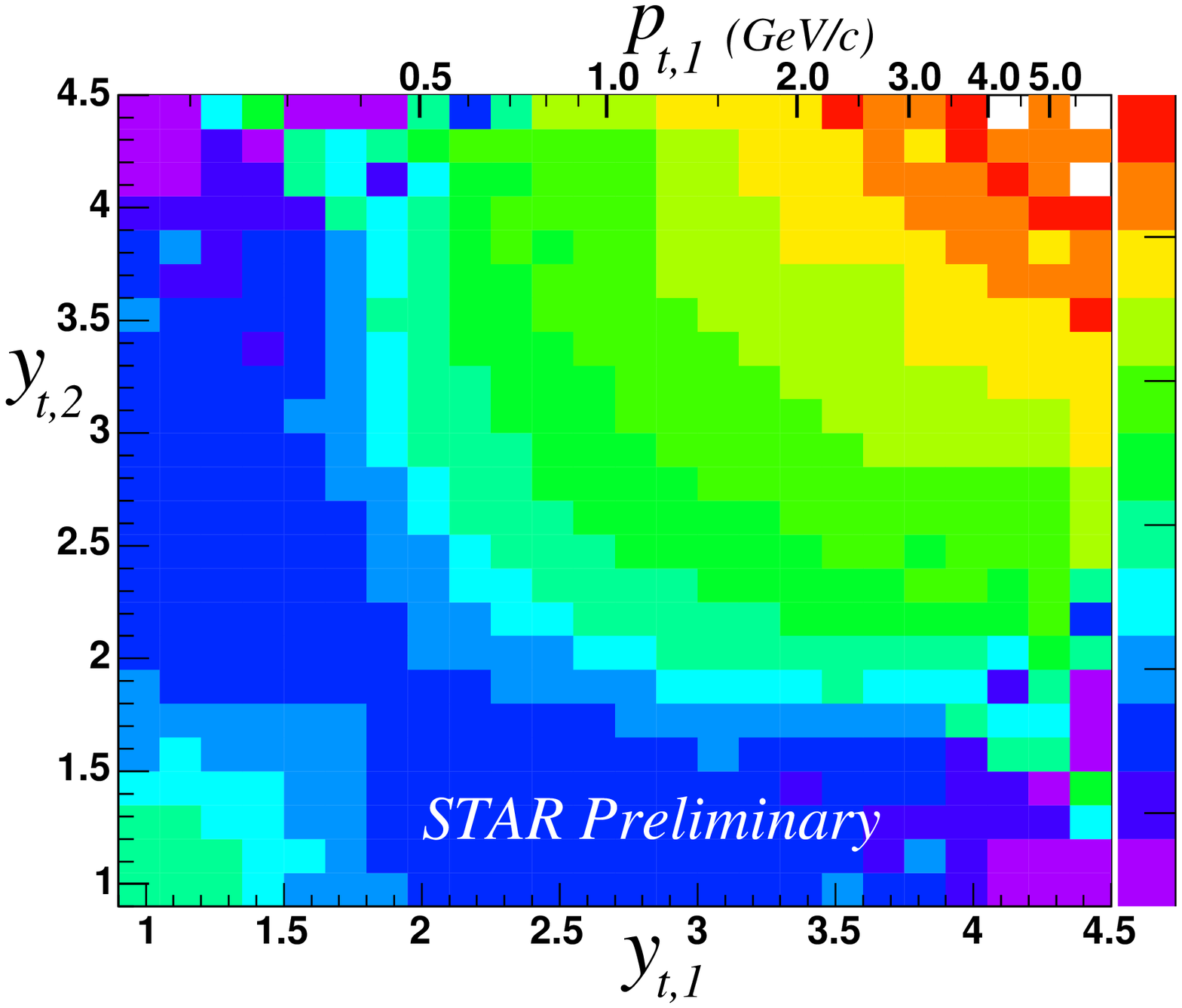}
\includegraphics[height=1.35in,width=1.63in]{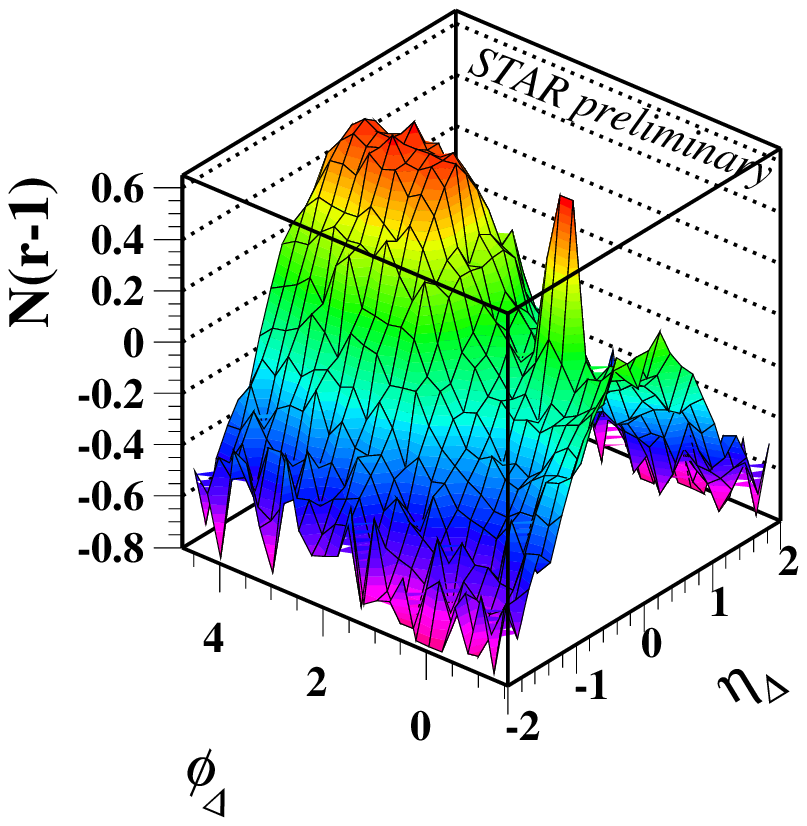}
\caption{\label{fig1}
Left:  preliminary distribution $\bar N (\hat r - 1)$ on $y_t \otimes y_t$ for the CI charge combination in the full STAR $(\eta,\phi)$ acceptance. Right: same quantity plotted as a joint autocorrelation on difference variables $( \eta_\Delta, \phi_\Delta)$, for the US soft component ($y_t < 2$ or $p_t < 0.5$ GeV/c).}
\end{figure}

In Fig.~\ref{fig1} (left panel) is plotted correlation measure $\bar N (\hat r - 1)$ for the CI charge combination on space $y_t \otimes y_t$. A $p_t$ scale is included at the top of the panel. 
The distribution has two components which can be separated by a cut at $y_t = 2$ corresponding to $p_t \sim 0.5$ GeV/c. The lower-left (soft) component is associated with longitudinal string fragmentation. The upper-right (hard) component is associated with jet fragments from minimum-bias (no trigger particle selected) hard parton scattering. Conventional jet analysis, invoking a high-$p_t$ trigger particle, typically defines an L-shaped region in the upper-right corner of Fig.~\ref{fig1} (left panel), bounded below by $y_t = 4$ ($p_t \sim 4$ GeV/c) and $y_t = 3.4$ ($p_t \sim 2$ GeV/c) and above by $y_t = 4.5$ ($p_t \sim 6$ GeV/c) for example~\cite{awayside}. In the present analysis we include all of the mimimum-bias parton fragments seen in Fig.~\ref{fig1} for $y_t > 2$.

In Fig.~\ref{fig1} (right panel) is the soft-component joint autocorrelation for the US charge combination (quantum or HBT correlations then do not appear). There are two major features: 1) a gaussian peak on $\eta_\Delta$ resulting from charge ordering along the collision axis $z$ during string fragmentation~\cite{lund} and expressed on pseudorapidity $\eta$ through longitudinal (Bjorken) expansion, and 2) a dip at $\phi_\Delta = 0$ reflecting transverse momentum conservation on azimuth for particles in close proximity on the string. The narrow peak at (0,0) is contamination from photon-conversion electron pairs. We did not parameterize this CI soft component in the present study.

\section{Axial CI correlations}

\begin{figure}[h]
\includegraphics[keepaspectratio,width=3.3in]{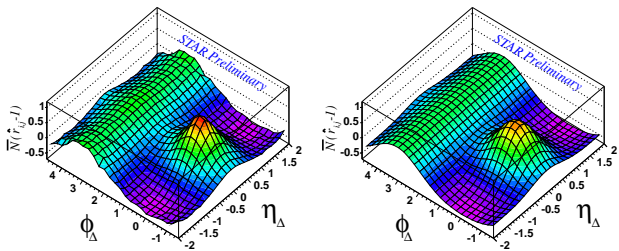}
\caption{\label{fig2}
Left: preliminary CI joint autocorrelation for the hard component, with $0.5 < p_t < 6$ GeV/c, showing minijet correlations. Right: model function fitted to data distribution.}
\end{figure}

Fig.~\ref{fig2} (left panel) shows the CI hard-component distribution ($0.5 < p_t < 6$ GeV/c) with two principal features: 1) a same-side peak ($|\phi_\Delta| < \pi/2$) representing minijet fragments from single partons, and 2) a broad away-side ridge ($|\phi_\Delta| > \pi/2$) representing correlations between particles from opposed pairs of partons. This correlation is extended on $\eta_\Delta$ because the center of momentum of parton pairs depends on the combination of momentum fractions $x$ for each parton collision and is broadly distributed relative to the laboratory reference frame.

Fig.~\ref{fig2} (right panel) shows a model function consisting of a 2D gaussian for the same-side peak and a single gaussian on $\phi_\Delta$ for the away-side ridge. Best-fit parameters include same-side widths $\sigma_{\eta_\Delta} = 0.502 \pm .004$, $\sigma_{\phi_\Delta} = .689 \pm .005$ and amplitude $=0.64 
\pm 0.01$, and away-side width $\sigma_{\phi_\Delta} = 1.05 \pm .01$ and amplitude $=0.38
\pm 0.01$. 
Because of the $y_t\otimes y_t$ cut definition those parameters correspond to a {\em minimum-bias} parton distribution. We can therefore describe the fragments as {\em mini}\,jets. This model parameterization is the correct p-p reference input for minimum-bias hard parton scattering in A-A collisions.


\section{Axial CD correlations}

\begin{figure}[h]
\includegraphics[keepaspectratio,width=3.3in]{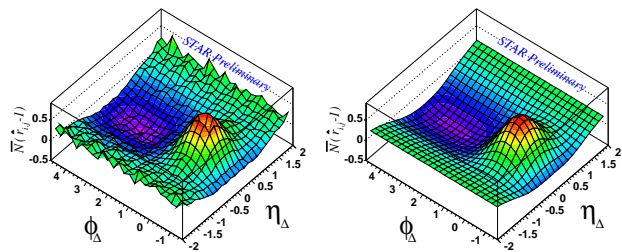}
\caption{\label{fig3}
Left: preliminary CD joint autocorrelation for the soft component, with $0.15 < p_t < 0.5$ GeV/c, showing net-charge correlations. Right: model function fitted to data distribution.}
\end{figure}

Fig.~\ref{fig3} (left panel) shows the soft-component joint autocorrelation for the CD charge combination with three features: 1) a negative 1D peak on $\eta_\Delta$ reflecting local suppression of net charge during string fragmentation, 2) a positive 2D peak at the origin representing quantum correlations (HBT) and 3) a small positive ridge beneath that peak reflecting 
greater canonical suppression of net transverse momentum on azimuth angle for unlike-sign pairs because they are closer on the string (related to the dip on azimuth difference noted in Fig.~\ref{fig1} right panel).

Fig.~\ref{fig3} (right panel) shows a model function consisting of a 1D gaussian on $\eta_\Delta$, a narrower 2D gaussian on $(\eta_\Delta,\phi_\Delta)$ centered at the origin and representing the HBT peak and a small-amplitude elongated 2D gaussian also centered at the origin. The 1D gaussian width is $\sigma_{\eta_\Delta} = 0.92 \pm 0.03$ with amplitude $= 0.75 \pm 0.2$. This 200 GeV result is, for $\sigma_{\eta_\Delta}$, comparable to a related 1D ISR analysis at $\sqrt{s} = 52.5$ GeV~\cite{isrpp}.

\begin{figure}[h]
\includegraphics[keepaspectratio,width=3.3in]{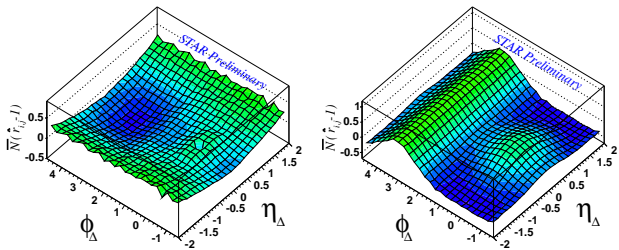}
\caption{\label{fig4}
Pythia V6.131, with soft CD (left) and hard CI (right) correlation components which compare well with data.}
\end{figure}

\section{Pythia Simulations}

Fig.~\ref{fig4} (left and right panels) shows respectively axial CD (soft) and axial CI (hard) components from the Pythia V6.131 Monte Carlo event generator which can be compared with corresponding data panels in Figs.~\ref{fig2} and \ref{fig3}. The general features of those distributions agree well with data. There are quantitative diffences which will be presented in a more detailed study. The CD gaussian peak width $\sigma_{\eta_\Delta}$ for Pythia at 200 GeV is $1.07 \pm 0.03$, whereas at 130 GeV it is $1.02 \pm 0.05$.

\section{Conclusions} 

The two-particle correlation structure of the hadron distribution from p-p collisions at 200 GeV can be separated into hard and soft components based on a simple cut at $y_t = 2$ ($p_t \sim 0.5$ GeV/c). The soft component is well represented by axial string fragmentation. The correlation structure is determined by local momentum and charge conservation which are manifested by charge-independent and charge-dependent structures. The hard component represents transverse parton fragmentation and also has charge-dependent (not considered here) and charge-independent manifestations. The jet-related CI angular correlations are determined by momentum conservation during initial hard scattering and fragmentation of the struck partons (or rather the excited color singlets which evolve from those partons).

The several two-particle data distributions have been represented by simple parameterized model functions. The model parameters reported in this short summary should be regarded as preliminary. In some cases comparison has been made to the Pythia Monte Carlo simulation and to data from the ISR at lower energies. These preliminary data and the Pythia Monte Carlo are in qualitative agreement, but there are significant quantitative differences that will be reported in a more comprehensive study. This minimum-bias representation of p-p correlation structure at 200 GeV is an essential reference for A-A collisions at RHIC.

\end{document}